# Algorithm for Evaluation of
# the Interval Power Function of Unconstrained Arguments

Evgueni Petrov

*A.P. Ershov Institute of Informatics Systems, Russian Academy of Science*
*prospekt akademika Lavrentjeva 6, Novosibirsk 630090*
*E-mail:* pes@iis.nsk.su

**Abstract**
We describe an algorithm for evaluation of the interval extension of the power function of variables $x$ and $y$ given by the expression $x^y$. Our algorithm reduces the general case to the case of non-negative bases.

## 1. Introduction

Despite the fact that algorithms for evaluation of the interval extension of the standard mathematical functions are well-known (performing arithmetic operations on intervals is even patented in the USA), there is no systematic approach to evaluation of the interval power function.

The existent software packages for interval calculations do calculate ranges of the power function restricted to certain domains. Most frequent constraints on the base $x$ and the exponent $y$ are $x \in \mathbf{R}$, $y \in \mathbf{Z}$ and $x \in \mathbf{R}_+ = [0, \infty)$, $y \in \mathbf{R}$. Such constraints complicate solution of problems with unknown exponents with the help of the existent interval software packages. Besides that, such constraints may interfere with symbolic transformations involving exponents if an interval software package and a symbolic preprocessor are used together. For example, reducing $x^3 =$ const to $x =$ const$^{1/3}$ is invalid for a negative const and a power function defined only for non-negative bases.

Most likely, such constraints on the base and the exponent emerged from developers' concern about discontinuous functions: the power function, as a function of two variables, is discontinuous or undefined for negative bases. The constraints on the base and the exponent actually specify domains of continuity for the power function.

In this paper we describe an algorithm for evaluation of the interval power function for arbitrary bases and exponents. We reduce evaluation in the general case to evaluation in the case of non-negative bases. Our algorithm is used by the UniCalc system [1].

The paper has the following structure. In Section 2 we reduce evaluation of the power function to the case of non-negative bases. In Section 3 we study our algorithm from Section 2 for some special cases. In Section 4 we discuss implementation of our algorithm for IEEE-754 and binary-decimal representations of real numbers. Section 5 concludes the paper.

## 2. Reduction to the Case of Non-Negative Bases

In this section we show that evaluation of the interval power function for arbitrary bases and exponents can be reduced to the case of non-negative bases, and give the corresponding algorithm.

Our key observation is that, in the standard topology of $\mathbf{R}^3$, the closure of the graph of the (full) power function is a union of a small number of parts that can be transformed, by reflections at the coordinate planes, into the graph of the power function restricted to non-negative bases. Thus the general case of evaluation of the interval power function is not much harder than the case of non-negative bases.

Below we present the detailed reasoning.

To begin with, we recall the domain of the power function. More or less elementary considerations from theory of functions of complex variables imply that, for a real base $x$ and a real exponent $y$, the power function is defined (takes a real value) if and only if one of the following conditions is satisfied:

- $y$ is an irreducible fraction with an odd denominator;
- $y$ is an irreducible fraction with an even denominator (and an odd numerator), $x \in \mathbf{R}_+$;
- $y$ is irrational, $x \in \mathbf{R}_+$.

Now we split the domain of the power function into such pieces that, in each of them, the power function is expressible in terms the restriction $pow_0$ of the power function to non-negative bases. For any subset $X \subseteq$

**R**, denote by eo($X$) (respectively oo($X$)) the subset of $X$ that consists of irreducible fractions whose numerator is even and denominator is odd (respectively, both are odd). Then, for a base $x$ and an exponent $y$, the power function is expressed in terms of $pow_0$ as follows:

$pow_0(x, y)$             if $x \in \mathbf{R}_+$;

$pow_0(-x, y)$           if $y \in eo(\mathbf{R})$, $x \notin \mathbf{R}_+$;

$-pow_0(-x, y)$         if $y \in oo(\mathbf{R})$, $x \notin \mathbf{R}_+$.

The graph of the power function is the union of the three parts specified by the above conditions. In the standard topology of $\mathbf{R}^3$, the closure of each of these parts can be obtained from the graph of $pow_0$ by reflections at the coordinate planes. The above decomposition can be inferred "automatically" from the conditions that define the domain of the power function by matching all possible forms of the exponent ("even/odd", "odd/odd", "odd/even", irrational) against the sign of the base.

Now we reduce the interval extension **pow** of the power function to the interval extension **pow₀** of the restriction $pow_0$ of the power function to non-negative bases. Recall that the interval extension **f** of any function $f$ of variables $x$ and $y$ maps any box (**x**, **y**) to the interval cch({$f(x, y) \mid x \in \mathbf{x}, y \in \mathbf{y}$}) where cch(·) denotes the convex hull of the closure of subsets of **R**.

Denote the sets {$pow_0(x, y) \mid x \in X, y \in Y$} and {$-x \mid x \in X$} by $pow_0(X, Y)$ and $-X$. According to our reduction of the power function to $pow_0$ and the definition of interval extensions, **pow**(**x**, **y**) = cch($pow_0$(**x**, **y**) ∪ $pow_0$(−**x**, eo(**y**)) ∪ −$pow_0$(−**x**, oo(**y**))) for any box (**x**, **y**). Note that cch($A \cup B$) = cch(cch($A$) ∪ cch($B$)). Thus **pow**(**x**, **y**) = cch( **pow₀**(**x**, **y**) ∪ cch($pow_0$(−**x**, eo(**y**))) ∪ −cch($pow_0$(−**x**, oo(**y**))) ).

It remains to express cch($pow_0$(−**x**, eo(**y**))) and cch($pow_0$(−**x**, oo(**y**))) in terms of **pow₀**.

For a non-singleton interval **y**, each of these terms equals cch($pow_0$(−**x**, **y**)). This conclusion follows from three facts. First, $pow_0$ is continuous. Second, continuous functions map closed sets to closed sets. Third, closure of eo(**y**) and oo(**y**) is **y** for a non-singleton interval **y**. Thus, for a non-singleton interval **y**, we get **pow**(**x**, **y**) = cch( **pow₀**(**x**, **y**) ∪ **pow₀**(−**x**, **y**) ∪ −**pow₀**(−**x**, **y**) ).

For a singleton interval **y**, there are 4 possibilities: to contain an irrational number, to contain a fraction of the form "even/odd", to contain a fraction of the form "odd/odd", and to contain a fraction of the form "odd/even". In each of these cases, the reductions in the expression for **pow**(**x**, **y**) are trivial (see below).

Summing up the above considerations, we get that the interval extension **pow** of the power function maps any box (**x**, **y**) to one of the following intervals:

- cch( **pow₀**(**x**, **y**) ∪ **pow₀**(−**x**, **y**) ∪ −**pow₀**(−**x**, **y**) )    if **y** is a non-singleton interval;
- cch( **pow₀**(**x**, **y**) ∪ **pow₀**(−**x**, **y**) )                   if **y** contains a single fraction of the form "even/odd";
- cch( **pow₀**(**x**, **y**) ∪ −**pow₀**(−**x**, **y**) )                if **y** contains a single fraction of the form "odd/odd";
- **pow₀**(**x**, **y**)                                            if **y** contains a single fraction of the form "odd/even";
- **pow₀**(**x**, **y**)                                            if **y** contains a single irrational number.

### 3. Study of Special Cases

In this section we study special cases that illustrate evaluation of the interval extension of the power function according the approach from Section 2.

**Special Case 3.1** (non-negative base).
If the base is non-negative, each of the final five cases from Section 2 is reduced to **pow₀**(**x**, **y**).

**Special Case 3.2** (negative base).
If the base is negative, the expression for **pow**(**x**, **y**) from Section 2 takes the following form:

- cch( **pow₀**(−**x**, **y**) ∪ −**pow₀**(−**x**, **y**) )          if **y** is a non-singleton interval;
- **pow₀**(−**x**, **y**)                                       if **y** contains a single fraction of the form "even/odd";
- −**pow₀**(−**x**, **y**)                                  if **y** contains a single fraction of the form "odd/odd".

We see that, if the base is negative and **y** is a non-singleton interval, the interval **pow**(**x**, **y**) is centered at zero. In other words, narrowing the interval exponent does help to determine the sign of the result of exponentiation, if the base is negative. This fact agrees with the fact that, for any negative number $x$ and any irreducible fraction $v$ of the form "odd/odd", there is an irreducible fraction $w$ of the form "even/odd" such that the absolute values of $x^v$ and $x^w$ are however close but their signs are opposite.

**Special Case 3.3** (intervals bounded by IEEE-754 floating point numbers).
Of practical interest is the restriction **pow_f** of the interval extension **pow** of the power function to the set of intervals bounded by floating point numbers in the IEEE-754 format [2]. Because the set of floating point numbers in the IEEE-754 format does not contain irrational numbers or irreducible fractions with odd denomi-nators different from 1 and –1, **pow**$_f$ maps any intervals **x** and **y** bounded by floating point numbers to one of the following intervals:

- cch(**pow**$_{0f}$(**x**, **y**)∪**pow**$_{0f}$(–**x**, **y**)∪–**pow**$_{0f}$(–**x**, **y**))   if **y** is a non-singleton interval;
- cch(**pow**$_{0f}$(**x**, **y**)∪**pow**$_{0f}$(–**x**, **y**))   if **y** contains a single even integral number;
- cch(**pow**$_{0f}$(**x**, **y**)∪–**pow**$_{0f}$(–**x**, **y**))   if **y** contains a single odd integral number;
- **pow**$_{0f}$(**x**, **y**)   if **y** contains a single floating point fraction.

Above **pow**$_{0f}$ denotes restriction of **pow**$_0$ to floating point intervals.

### 4. Remarks on Computer Implementation of Our Approach

In this section we comment on computer implementation of the approach proposed in Section 2.

We already noted in Section 1 that there exists a number of software packages for interval computations that calculate interval extension of the power function for non-negative bases. Therefore reduction of the general case to the case of non-negative bases is reasonable.

From a developer's point of view, the reduction to the case of non-negative bases from Section 2 is a trivial chain of if-then-else statements. However, the amount of calculations that are needed to check the conditions in those if-then-else statements depends on the format used for storing real numbers.

If interval bounds are stored in the IEEE-754 format, then, as we saw in Special Case 3.3, most CPU time is spent to calculate **pow**$_{0f}$(**x**, **y**) (we use the notation from Section 3). Thus, in the case of IEEE-754 floating point numbers, removing constraints on the base and the exponent just doubles the amount of calculations. Such payment seems acceptable.

If interval bounds are stored in a format that permits a complicated set of irreducible fractions with odd denominators, then checking the conditions on singleton interval exponents may require much more CPU time. An example of such an "inconvenient" (from the point of view of Section 2) floating point format is the binary-decimal format.

Finally we note that the overestimation introduced by **pow**$_f$ is determined by the overestimation introduced by **pow**$_{0f}$. If **pow**$_{0f}$ does not overestimate ranges of the power function for non-negative bases (which is unlikely in general), then the function **pow**$_f$ does neither. A correct implementation of the approach from Section 2 guarantees only that **pow**$_f$(**x**, **y**) contains the true interval **pow**(**x**, **y**).

### 5. Conclusion

In our paper we described an algorithm for evaluation of the interval extension of the power function for arbitrary intervals. We showed that evaluation in the general case can be reduced to the case of non-negative base. Our key observation is that the closure of the graph of the power function is a union of a small number of pieces that can be transformed into the graph of the power function restricted to non-negative bases.

Computer implementation of our approach depends on the format used for storing floating point numbers. For the popular IEEE-754 format, our approach to removing constraints on the base and the exponent just doubles the amount of calculations compared to the case of non-negative bases.

The author thanks A.P. Ershov Institute of Informatics Systems for financial support of this research.